\documentclass[aps, twocolumn, superscriptaddress, showpacs, nofootinbib, longbibliography]{revtex4-1}

\usepackage[utf8]{inputenc}
\usepackage[T1]{fontenc}
\usepackage{ae,aecompl} 
\usepackage{graphicx}
\graphicspath{{figures/}} % Directory in which figures are stored

\usepackage{amsmath}
\usepackage{color}
\usepackage{amssymb}
\usepackage{latexsym}
\usepackage{wasysym}
\usepackage{psfrag}
\usepackage{ifthen}
\usepackage[citecolor=blue,colorlinks=true]{hyperref}
\usepackage{longtable}
\usepackage{float}
\usepackage[utf8]{inputenc}
\usepackage{lineno}
\usepackage{units}
\usepackage[title]{appendix}
\usepackage{aas_macros}
\usepackage{ulem}
\usepackage{subfigure}
\usepackage{multirow}
\def\mnras{Mon. Not. Roy. Astron. Soc.}
\def\apjl{Astrophys. J. Lett.}
\def\apj{Astrophys. J.}
\def\prd{Phys.~Rev.~D}
\def\aap{A\&A}

\definecolor{maob}{rgb}{0.6,0.1,0.2}

\newcommand{\Msol}{M_{\odot}}

\begin{document}
%\linenumbers

\title{Inference of proto-neutron star properties from gravitational-wave data in core-collapse supernovae}

\author{Marie-Anne~Bizouard}
\affiliation{Artemis, Universit\'e C\^ote d'Azur, Observatoire C\^ote d'Azur, CNRS, CS 34229, F-06304 Nice Cedex 4, France}

\author{Patricio~Maturana-Russel}
\affiliation{Department of Statistics, The University of Auckland, Auckland, New Zealand}
\affiliation{Department of Mathematical Sciences, Auckland University of Technology, Auckland, New Zealand}

\author{Alejandro~Torres-Forn\'e}
\affiliation{Max Planck Institute for Gravitationalphysik (Albert Einstein Institute), D-14476 Potsdam-Golm, Germany}
\affiliation{Departamento de Astronom\'{\i }a y Astrof\'{\i }sica, Universitat de Val\`encia, E-46100 Burjassot, Val\`encia, Spain}

\author{Martin~Obergaulinger}
\affiliation{Departamento de Astronom\'{\i }a y Astrof\'{\i }sica, Universitat de Val\`encia, E-46100 Burjassot, Val\`encia, Spain}

\author{Pablo~Cerd\'a-Dur\'an}
\affiliation{Departamento de Astronom\'{\i }a y Astrof\'{\i }sica, Universitat de Val\`encia, E-46100 Burjassot, Val\`encia, Spain}

\author{Nelson~Christensen}
\affiliation{Artemis, Universit\'e C\^ote d'Azur, Observatoire C\^ote d'Azur, CNRS, CS 34229, F-06304 Nice Cedex 4, France}
\affiliation{Physics and Astronomy, Carleton College, Northfield, MN 55057, USA}

\author{Jos\'e~A.~Font}
\affiliation{Departamento de Astronom\'{\i }a y Astrof\'{\i }sica, Universitat de Val\`encia, E-46100 Burjassot, Val\`encia, Spain}
\affiliation{Observatori Astron\`omic, Universitat de Val\`encia, E-46980, Paterna, Val\`encia, Spain} 

\author{Renate~Meyer}
\affiliation{Department of Statistics, The University of Auckland, Auckland, New Zealand}

\begin{abstract}
The eventual detection of gravitational waves from core-collapse supernovae (CCSN) will help improve our current understanding of the explosion mechanism of massive stars. The stochastic nature of the late post-bounce gravitational wave signal due to the non-linear dynamics of the matter involved and the large number of degrees of freedom of the phenomenon make the source parameter inference problem very challenging. In this paper we take a step towards that goal and present a parameter estimation approach which is based on the gravitational waves associated with oscillations of proto-neutron stars (PNS). Numerical simulations of CCSN have shown that buoyancy-driven g-modes are responsible for a significant fraction of the gravitational wave signal and their time-frequency evolution is linked to the physical properties of the compact remnant through universal relations, as demonstrated in~\citep{Torres:2019b}. We use a set of 1D CCSN simulations to build a model that relates the  evolution of the PNS properties with the frequency of the dominant g-mode, which is extracted from the gravitational-wave data using a new algorithm we have developed for our study. The model is used to infer the time evolution of a combination of the mass and the radius of the PNS. The performance of the method is estimated employing simulations of 2D CCSN waveforms covering a progenitor mass range between 11 and 40 solar masses and different equations of state. Considering signals embedded in Gaussian gravitational wave detector noise, we show that it is possible to infer PNS properties for a galactic source using Advanced LIGO and Advanced Virgo data at design sensitivities. Third generation detectors such as Einstein Telescope and Cosmic Explorer will allow to test distances of ${\cal O}(100\, {\rm kpc})$. %\tf{Caveats? Gaussian noise?}

\end{abstract}

\maketitle

% !TEX root = ccsn.tex

\section{Introduction}

The life of sufficiently massive stars, i.e.~those born with masses between $\sim 8$~M$_\odot$ and $\sim 120$~M$_\odot$, ends with the collapse of {the} iron core under {its} own gravity, leading {to} the formation of a neutron star {(NS)} or a black hole (BH), {and} followed (typically but not necessarily in the BH case) by {a supernova} explosion.
Nearby core-collapse supernova (CCSN) explosions are expected to be sources of gravitational waves (GWs) and they could be the next great discovery of current ground-based observatories. However, these are relative rare events. A neutrino-driven explosion~\citep{Bethe:1990} is the most likely outcome in the case of slowly rotating cores, which are present in the bulk of CCSN progenitors.
The emitted GWs could be detected with the advanced ground-based GW detector network, Advanced LIGO (aLIGO)~\citep{TheLIGOScientific:2014jea}, Advanced Virgo (AdV)~\citep{TheVirgo:2014hva} and KAGRA~\citep{Aso:2013eba}, within $\sim$ \unit[5]{kpc}~\citep{Gossan:2016,TargetedSNSearchO12}. Such a galactic event has a rate of about $2-3$ per century~\citep{Adams:2013,Rozwadowska:2021}.
For the case of rapidly rotating progenitor cores the result is likely a magneto-rotational explosion, yielding  a more powerful GW signal that could be detected within \unit[50]{kpc} and, for some extreme models, up to \unit[5--30]{Mpc}~\citep{Gossan:2016,TargetedSNSearchO12}. However, only about $1\%$ of the electromagnetically observed events show signatures of fast rotation (broad-lined type Ic SNe~\citep{Li:2011b} or events associated with long GRBs~\citep{Chapman:2007}) making this possibility a subdominant channel of detection with a galactic event rate of $\sim$ \unit[$10^{-4}$]{$\rm {yr}^{-1}$}.
Despite the low rates, CCSN are of great scientific interest because they produce complex GW signals which could provide significant clues about the physical processes at work after the gravitational collapse of stellar cores.

In the last decade significant progress has been made in the development of numerical codes, {in particular in the treatment of multidimensional effects~\citep{BMueller:2020}.} In the case of  neutrino-driven explosions, the GW emission is {primarly induced by instabilities developed at the newly formed proto-neutron star (PNS) and by the non-spherical accreting flow of hot matter over its surface~\citep{Kotake:2017}.  These} dynamics excite the different modes of oscillation of the PNS which ultimately leads to the emission of GWs. The frequency and time evolution of these modes carry information about the properties of the GW emitter and could allow to perform PNS asteroseismology.

%{[PCD: I would remove the discussion on the inference of the progenitor properties (commented now). ]}
%Unluckily, in this case it is not posible to relate the GW emission with the properties (mass, rotation rate, metallicity or magnetic fields) of the progenitor stars.  A large number of physical processes are involved and their role is not completely understood. For instance,  uncertainties in the stellar evolution models of massive stars or in the nuclear and weak interactions necessary for the equation of state (EoS) of nuclear matter or the neutrino interactions. Furthermore, the stochastic and chaotic nature of the instabilities is transferred to the GW emission, resulting in the same progenitor leading significantly different waveforms.
%The large number of physical ingredients in addition to the necessary accuracy of the modelling of complex multidimensional interactions requires large computational resources. One simulation of a single progenitor explosion  in 3D with accurate neutrino transport and realistic EoS can take several months of intense calculations on a scientific supercomputer facility. This complicates the systematic exploration of the progenitor parameters.

All multidimensional numerical simulations show the systematic appearance in time-frequency diagrams (or spectrograms) of a distinct and relatively narrow feature 
%
%{The main feature appearing systematically in the GW spectrum of multidimensional numerical simulations is a strong and relatively narrow \mab{oscillation} 
%
during the post-bounce evolution of the system, with frequency rising 
from about \unit[100]{Hz} up to a few kHz (at most) and a typical duration of \unit[0.5 -- 1]{s}. This feature has been interpreted as a continuously excited gravity mode (g-mode, see~\citep{kokkotas,Friedman:2013} for a definition in this context) of the PNS~\citep{murphy:09, mueller:13gw, Cerda:2013, Yakunin:2015wra, Kuroda:2016, Andresen:2017}. 
In these models the monotonic raise of the frequency of the mode is related to the contraction of the PNS. The {typical} frequencies of {these} modes make them interesting targets for ground-based GW interferometers. 
 
 The {properties of} g-modes in hot {PNSs} have been studied since the 1990s {by means of linear perturbation analysis of background PNS models}. The oscillation modes connected with the surface of hot PNSs were first considered by McDermott {\it et al.}~\citep{McDermott:1983}. Additionally, the stratified structure of the PNS allows for the presence of different types of g-modes related to the fluid core~\citep{Reisenegger:1992}. Many subsequent works used simplified neutron star models assuming an equilibrium configuration {as a background}, to study the effect of rotation~\citep{Ferrari:2004}, general relativity \citep{Passamonti:2005}, non-linearities~\citep{Dimmelmeier:2006}, phase transitions~\citep{Kruger:2015} and realistic equation of state~\citep{Camelio:2017}. {Only recently, there have been efforts to incorporate more suitable backgrounds based on numerical simulations in the computation of the mode structure and evolution~\citep{Sotani:2016,Torres:2018, Morozova:2018, Torres:2019a,Torres:2019b,Sotani:2019,WS:2019,Sotani:2020a, Sotani:2020b}}.
 
Using results from 2D CCSN numerical simulations as a background~\citep{Torres:2018, Torres:2019a} found that the eigenmode spectrum of the region within the shock (including the PNS and the post-shock region) 
shows a good match to the mode frequencies and to the features observed in the GW spectrum of the same simulations (specially when space-time perturbations are included~\citep{Torres:2019a}).

This reveals that it is posible to perform CCSN asteroseismology {under realistic conditions} and serves as a starting point to carry out inference of astrophysical parameters of PNSs. Further work was presented in~\citep{Torres:2019b} who found that it is possible to derive simple relations between the instantaneous frequency of the g-mode and the mass and radius of the PNS {at each time of the numerical evolutions}. These relations are universal as they do not depend {on the equation of state (EOS) or on the mass of the progenitor {and they only depend weakly on} the numerical code used {(see discussion in Section~\ref{sec:simulations})}. {Similar universal relations have been discussed by~\citep{Sotani:2020a,Sotani:2020b} who also found that they do not depend on the dimensionality (1D, 2D or 3D) of the numerical simulation used as a background.

Previous data analysis efforts have focused on the reconstruction of the GW strain amplitude
without assuming a particular signal model~\citep{Summerscales:2008,Klimenko:2015ypf,CornLitt}. As the amount of numerical
simulations increased other methods using waveform catalogs have been proposed to identify
the supernova explosion mechanism.
Among them, principal component analysis helps at reducing the complexity of CCSN waveforms to fewer
parameters~\citep{Heng:2009,roever:09,Edwards:2014,powell:2016,powell:2017,Roma:2019kcd}.
The classification challenge is also well addressed with deep learning methods~\citep{astone:2018,Chan:2019fuz}.
    
{In this work we introduce a method to infer PNS properties, namely a combination of the mass and radius, using GW information. For this purpose we have developed an algorithm to  
extract the time-frequency evolution of the main feature in the spectrograms of the GW emission of 2D simulations of CCSN. This feature corresponds to the $^2\rm{g}_2$ mode, according to the nomenclature used in~\citep{Torres:2019b} (different authors may have slightly different naming convention). Next, we use the universal relations obtained by~\citep{Torres:2019b}{, based on a set of 1D simulations,} to infer the time evolution of the ratio $M_{\rm PNS}/R_{\rm PNS}^2$ (the PNS surface gravity), where  $M_{\rm PNS}$ and $R_{\rm PNS}$ are the mass and the radius of the PNS.} Finally, using 2D CCSN waveform corresponding to different progenitor masses we estimate the performance of the algorithm for current and future generation of ground-based GW detectors.

This paper is organized as follows. Section II provides details of the CCSN simulations used in our work. The algorithm that we employ to extract the time evolution of the PNS surface gravity is discussed in Section III. Section IV shows the performance of our inference method and presents our main results. Finally, our findings are summarized in Section V. Appendix A discusses specific details related to the reconstruction of the g-mode.

\bigskip

% !TEX root = ccsn.tex
\section{Core collapse supernova simulations}
\label{sec:simulations}

Unlike other methods used in GW astronomy the algorithm proposed in {this work} does not require accurate
waveforms {in order to infer the properties of the PNS.} Instead, it relies on the evolution of the oscillation
frequency of some particular modes visible in the GW spectrum.
The frequency of these modes depends in a universal way on the surface gravity of the PNS, $r\equiv M_{\rm PNS}/R_{\rm PNS}^2$~\citep{Torres:2019b}. Therefore, if at a given time GW emission is observed at a certain
frequency $f$ then the value of the surface gravity can be determined unequivocally, within a certain error,
regardless of the details of the numerical simulation. 

In this work we use two sets of simulations: i) the {\it model set}, composed by 1D simulations, which is used to build the 
universal relation (model), $r(f)$, linking the ratio $r$ with the observed frequency $f$, and ii) the {\it test set}, composed by
2D simulations, 
for which we know both the GW signal and the evolution of the ratio, $r (t)$, and that is used to test
performance of the algorithm.

%{We have used two different numerical codes in our numerical simulations.} 
%CoCoNuT
%(one-dimensional models) and AENUS-ALCAR
%\citep{Just_et_al__2015__mnras__Anewmultidimensionalenergy-dependenttwo-momenttransportcodeforneutrino-hydrodynamics}
%(one- and two-dimensional models). 
%CoCoNuT
%\citep{Dimmelmeier:2002,Dimmelmeier:2005} is a code for general
%relativistic hydrodynamics coupled to the Fast Multigroup Transport
%scheme \citep{Mueller_Janka_2015_FMT} providing an approximate
%description of the emission and transport of neutrinos.
Both the {\it model set} and {\it test set} simulations have been generated using the AENUS-ALCAR code~\citep{Just:2015}
which combines special relativistic (magneto-)hydrodynamics, a modified Newtonian gravitational potential approximating the effects of general relativity~\citep{Marek_etal__2006__AA__TOV-potential}, and a spectral
two-moment neutrino transport solver~\citep{Just:2015}.
All siimulations include the relevant reactions between matter and neutrinos of all
flavours, i.e., emission and absorption by nucleons and nuclei,
electron-positron pair annihilation, nucleonic bremsstrahlung, and
scattering off nucleons, nuclei, and electrons.

For the {\it model set}, we use the $18$ spherically symmetric (1D) simulations of~\citep{Torres:2019a} including progenitors with zero-age main sequence (ZAMS) masses in the range $M_{\mathrm{ZAMS}}=$ \unit[11.2 -- 75]{$\Msol$}. The set contains simulations using six different EOS. Details can be found in~\citep{Torres:2019a}. The reason to use 1D simulations for the model set is that their computational cost is significantly smaller than that of multidimensional simulations which allows us to accumulate the statistics necessary to build a good model for $r(f)$.
For each time of each simulation we compute the ratio $r$ and the frequency of the $^2g_2$ mode by means of the linear analysis described in \cite{Torres:2018,Torres:2019a,Torres:2019b}.
 
 \begin{table}
 \centering
 \begin{tabular}{c|ccc|ccc}
  \hline
  Model & $M_\mathrm{ZAMS} $ & progenitor& EOS & $t_{\mathrm{f}}$& $t_{\rm explosion}$ & $M_{\mathrm{PNS, f}}$\\
  name& $[\Msol]$ & model & & $[\mathrm{s}]$& & $[\Msol]$ 
  \\ 
  \hline
  \texttt{s11} & 11.2 & \cite{Woosley_Heger_Weaver__2002__ReviewsofModernPhysics__The_evolution_and_explosion_of_massive_stars}& LS220 & 1.86 & $\times$ & 1.47 
  \\ 
  \texttt{s15} & 15.0 & \cite{Woosley_Heger_Weaver__2002__ReviewsofModernPhysics__The_evolution_and_explosion_of_massive_stars}& LS220 & 1.66 & $\times$ & 2.00 
    \\ 
  \texttt{s15S} & 15.0 & \cite{Woosley_Heger_Weaver__2002__ReviewsofModernPhysics__The_evolution_and_explosion_of_massive_stars}& SFHo & 1.75 & $\times$ & 2.02 
    \\ 
  \texttt{s15G} & 15.0 & \cite{Woosley_Heger_Weaver__2002__ReviewsofModernPhysics__The_evolution_and_explosion_of_massive_stars}& GShen & 0.97 & $\times$ & 1.86
     \\ 
  \texttt{s20} & 20.0 & \cite{Woosley_Heger_Weaver__2002__ReviewsofModernPhysics__The_evolution_and_explosion_of_massive_stars}& LS220 & 1.53 & $\times$ & 1.75 
    \\ 
  \texttt{s20S} & 20.0 & \cite{Woosley_Heger__2007__physrep__Nucleosynthesisandremnantsinmassivestarsofsolarmetallicity} & SFHo & 0.87 & $\times$ & 2.05 
  \\ 
  \texttt{s25} & 25.0 & \cite{Woosley_Heger_Weaver__2002__ReviewsofModernPhysics__The_evolution_and_explosion_of_massive_stars}& LS220 & 1.60 & $0.91$ & 2.33 
    \\ 
  \texttt{s40} & 40.0 & \cite{Woosley_Heger_Weaver__2002__ReviewsofModernPhysics__The_evolution_and_explosion_of_massive_stars}& LS220 & 1.70 & $1.52$ & 2.23 
    \\ \hline
 \end{tabular}
 \caption{%%
  List of axisymmetric simulations {used for the {\it test set}}. 
  {The last three columns show, the post-bounce time at the end of the
  simulation, the one at the onset of the explosion (non exploding models marked
  with $\times$), and the PNS mass at the end of the simulation.}
 }
 \label{Tab:2dSimList}
\end{table}

{For the {\it test set}, we use $8$ axisymmetric (2D) simulations performed with the {\sc AENUS-ALCAR} code
(see Table~\ref{Tab:2dSimList} for a list of models).
All of these simulations but model \texttt{s20S} use a selection of progenitors with masses in the range} $M_{\mathrm{ZAMS}}=$ \unit[11.2 -- 40]{$\Msol$}
 evolved through the hydrostatic phases by~\citep{Woosley_Heger_Weaver__2002__ReviewsofModernPhysics__The_evolution_and_explosion_of_massive_stars}.
We performed one simulation of each stellar model using the EOS of~\citep{Lattimer_Swesty__1991__NuclearPhysicsA__LS-EOS} with an incompressibility of $K=$ \unit[220]{MeV} (LS220) and added comparison simulations with the SFHo EOS~\citep{Steiner_et_al__2013__apj__Core-collapseSupernovaEquationsofStateBasedonNeutronStarObservations} and the GShen EOS~\citep{Shen_et_al__2011__prc__Newequationofstateforastrophysicalsimulations} for the {progenitor} with $M_{\mathrm{ZAMS}}=$ \unit[15]{$\Msol$}.
To this set of simulations we add the waveform of a 2D model used in~\citep{Torres:2019a}, denoted \texttt{s20S}. It corresponds to a star with the same initial mass, $M_{\mathrm{ZAMS}}=$ \unit[20]{$\Msol$}, as for one of the other seven axisymmetric simulations, but was taken from a newer set of stellar-evolution models~\citep{Woosley_Heger__2007__physrep__Nucleosynthesisandremnantsinmassivestarsofsolarmetallicity}. It was evolved with the SFHo EOS.

{For all the simulations,} we mapped the pre-collapse state of the stars to a spherical
coordinate system with $n_r = 400$ zones in the radial direction
distributed logarithmically with a minimum grid width of
$(\Delta r)_{\mathrm{min}}=$ \unit[400]{m} and an outer radius of
$r_{\mathrm{max}} =$ \unit[$8.3 \times 10^{9}$]{cm} and
$n_{\theta} = 128$ equidistant cells in the angular (polar) direction. For the
neutrino energies, we used a logarithmic grid with $n_e = 10$ bins up
to \unit[240]{MeV}.
{Unlike the model set, the simulations in the test set are not 1D because we need to 
extract the GW signal, which is a multi-dimensional effect. For each simulation
the GW signal, $h_+(t)$, is extracted by means of the quadrupole formula and we compute the 
time evolution of the surface gravity, $r(t)$.}

All spherical and most axisymmetric models we evolved fail to achieve shock
revival during the time of our simulations. Only the two stars with
the highest masses, \texttt{s25} and \texttt{s40}, develop relatively
late explosions in axisymmetry. Consequently, mass accretion onto the
PNSs proceeds at high rates for a long time in all cases and causes
them to oscillate with their characteristic frequencies. The final
masses of the PNSs are in the range of
$M_{\mathrm{PNS}} =$ \unit[1.47 -- 2.33]{$\Msol$}, i.e., likely insufficient for
producing a black hole.

%%% Local Variables:
%%% TeX-master: "ccsn"
%%% End:
\bigskip

% !TEX root = ccsn.tex
\section{Description of the method}
\label{methods}

\begin{figure}[t]
 \centering
 \includegraphics[width=0.45\textwidth]{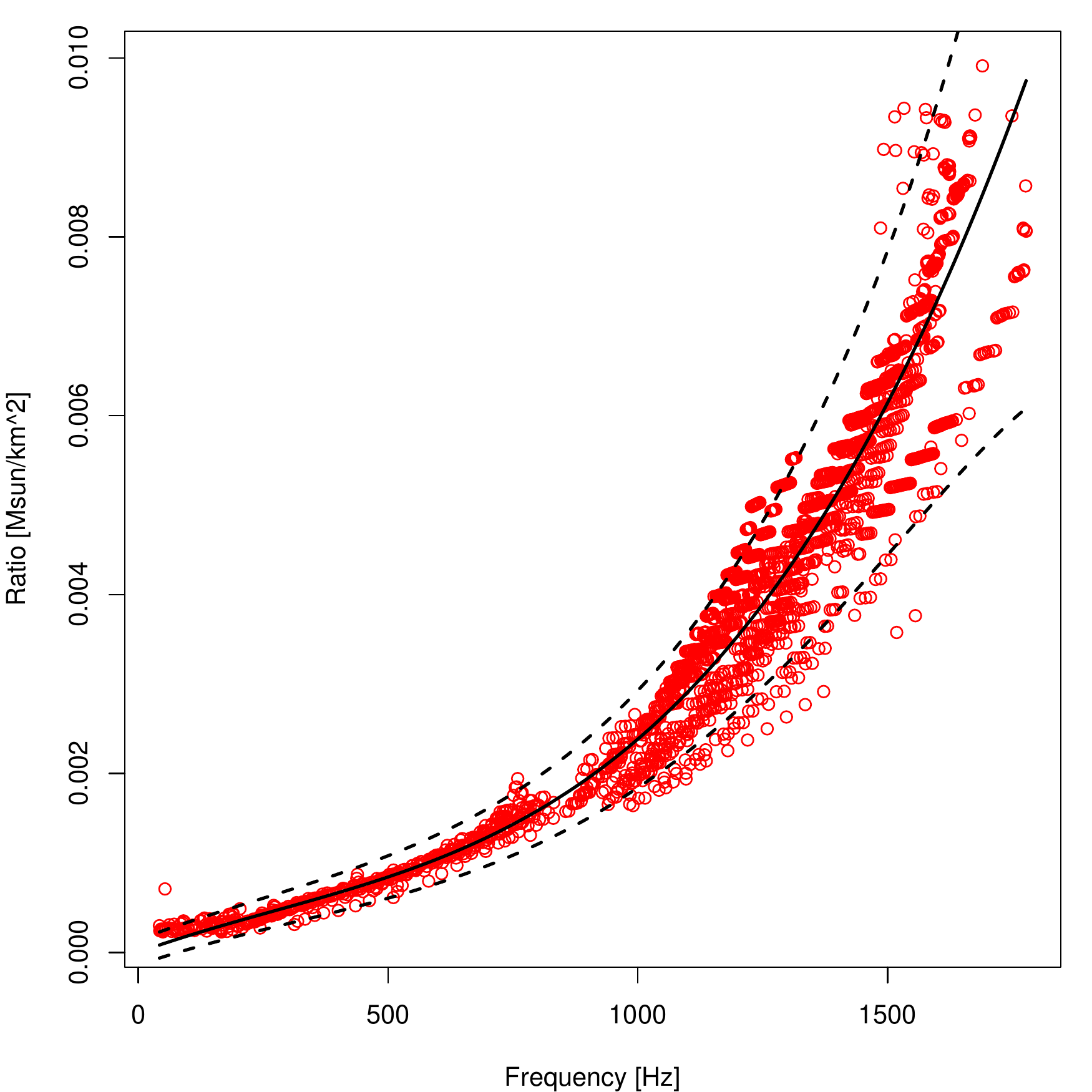}
 \caption{Ratio $M_{\rm PNS}/R_{\rm PNS}^2$ from our 18 1D simulations of the model set. The solid line is the maximum likelihood estimate of heteroscedastic cubic model with 95\% confidence bands (dashed lines) considering the 18 simulation data points.} 
 \label{fig:LMVAR}
\end{figure}

We next outline our strategy for estimating the time evolution of $r(t)$
%ratio $r=M_{\rm PNS}/R_{\rm PNS}^2$ (in units of solar mass and km) 
from the observation of the $\mbox{}^2g_2$ oscillation mode in the GW detector data.
%An integral part of this strategy is the universal relations that relate the
%characteristic frequency of the PNS oscillation $f$, $g$ and $p$ modes with the mass
%and the radius of the PNS, the shock radius and the total mass inside the shock as
%demonstrated in \cite{Torres:2019b}.
To build the model of the ratio $r$ as a function of the frequency $f$ we use the 
1D simulations of the {\it model set}. Figure~\ref{fig:LMVAR}
shows the data for the $18$ numerical simulations. 
%As identified by \cite{Torres:2019b}, the only systematic deviation from a single universal relation is the numerical code used in the simulations. 
%To avoid any systematic effect, we only use the $18$ simulations performed with the {\sc AENUS-ALCAR}
%code, which is the same code that was used in our test set.
%\mab{We defer a discussion of the possible consequences of this choice in Section \ref{sec:conclusion}.}
%The consequences of this choice are discussed in the conclusions.
Using these data we parametrize the discretized ratio $r_i$ with a cubic polynomial
regression with heteroscedastic errors
\begin{equation}
\label{eq:model1}
r_i=\beta_1 f_i + \beta_2 f_i^2 +\beta_3 f_i^3 + \epsilon_i\,,
\end{equation}
where $\epsilon_i$ are assumed to be independent zero-mean Gaussian errors with
variances $\sigma_i^2$ that increase with frequency $f_i$. The model for frequency-dependent
variances is
\begin{equation}
\log \sigma_i=\alpha_0+ \alpha_1 f_i + \alpha_2 f_i^2 + \delta_i\,,
\end{equation}
with independent and identically zero-mean Gaussian errors $\delta_i$. The R-package \texttt{lmvar}~\citep{lmvar:2019} that implements a maximum likelihood approach was used to fit the model.

The best fitting model amongst polynomials of degree 1, 2, and 3  was chosen according to
the Akaike information criterion with coefficients given in Table \ref{tab:model}, which is actually the model defined in \eqref{eq:model1}.  The data and fit of the model including 95\% confidence bands are displayed in
Figure~\ref{fig:LMVAR}.

%\begin{equation}\label{eq:universal}
%r_i = \beta_1 f_i + \beta_3 f_i^3 + \epsilon_i
%\end{equation}

%The best-fitting model achieves a coefficient of determination of $R^2=0.9812$.
%The data and fit of the model including 95\% confidence bands are displayed in
%Figure~\ref{fig:LMVAR}.

\begin{table}[h]
%  \begin{tabular}{lll}
%    \hline
%    Coefficient & Estimate & standard error \\
%    \hline
%    $\beta_1$   & $6.09 \times 10^{-7}$ & $1.75 \times 10^{-8}$ \\
%    $\beta_3$   & $6.24 \times 10^{-13}$ & $8.79 \times 10^{-15}$ \\
%    \hline
%  \end{tabular}

  \begin{tabular}{crr}
    \hline
    Coefficient & \multicolumn{1}{c}{Estimate} & Standard error \\
    \hline
   $\beta_1$  &  $ 2.00 \times 10^{-06}$ & $4.23 \times 10^{-08}$ \\   
   $\beta_2$  &  $-1.64 \times 10^{-9}$ & $9.99 \times 10^{-11}$ \\
   $\beta_3$  &  $ 2.03 \times 10^{-12}$ & $5.41 \times 10^{-14}$ \\
   $\alpha_0$ &  $-9.54 \times 10^{+00}$ & $6.80 \times 10^{-02}$ \\
   $\alpha_1$ &  $ 7.24 \times 10^{-04}$ & $1.56 \times 10^{-04}$ \\
   $\alpha_2$ &  $ 6.23 \times 10^{-07}$ & $8.15 \times 10^{-08}$ \\   
    \hline
  \end{tabular}
\caption{Estimate and standard error of the coefficients of the best fit model describing the ratio $r=M_{\rm PNS}/R_{\rm PNS}^2$ as function of the frequency of the $\mbox{}^2g_2$ mode.}\label{tab:model}
\end{table}

{We use this model to infer the properties of the simulations in the 
{\it test set} discussed in Section \ref{sec:simulations}.}
To describe the method we focus on the GW signal
of model {\texttt s20S}, originally
sampled at \unit[16384]{Hz} but downsampled at \unit[4096]{Hz}.
A spectrogram of this signal is shown in Figure \ref{fig:spectrogram} based on
autoregressive estimates~\citep{BrockwellPeterJ1991Tsta} of the local spectra for successive time intervals of length 200 with a 90\% overlap.
The dominant emission mode corresponds to the PNS oscillation $\mbox{}^2 g_2$-mode. We have
developed a time-frequency method to track the ridge $m(t) $ in the spectrogram,
taking into account that it is monotonically increasing with time. 
This is a property of the $\mbox{}^2 g_2$-mode whose frequency   
increases as the PNS becomes more massive and compact.
Starting from either the left- or right-most column of the time-frequency matrix
we identify and trace the sequence of amplitude peaks within a certain frequency
band given the monotonicity constraint. Specific details about the reconstruction of the $\mbox{}^2 g_2$-mode ridge 
are provided in Appendix~\ref{app:gmode}. 

\begin{figure}
 \centering
 \includegraphics[width=0.5\textwidth]{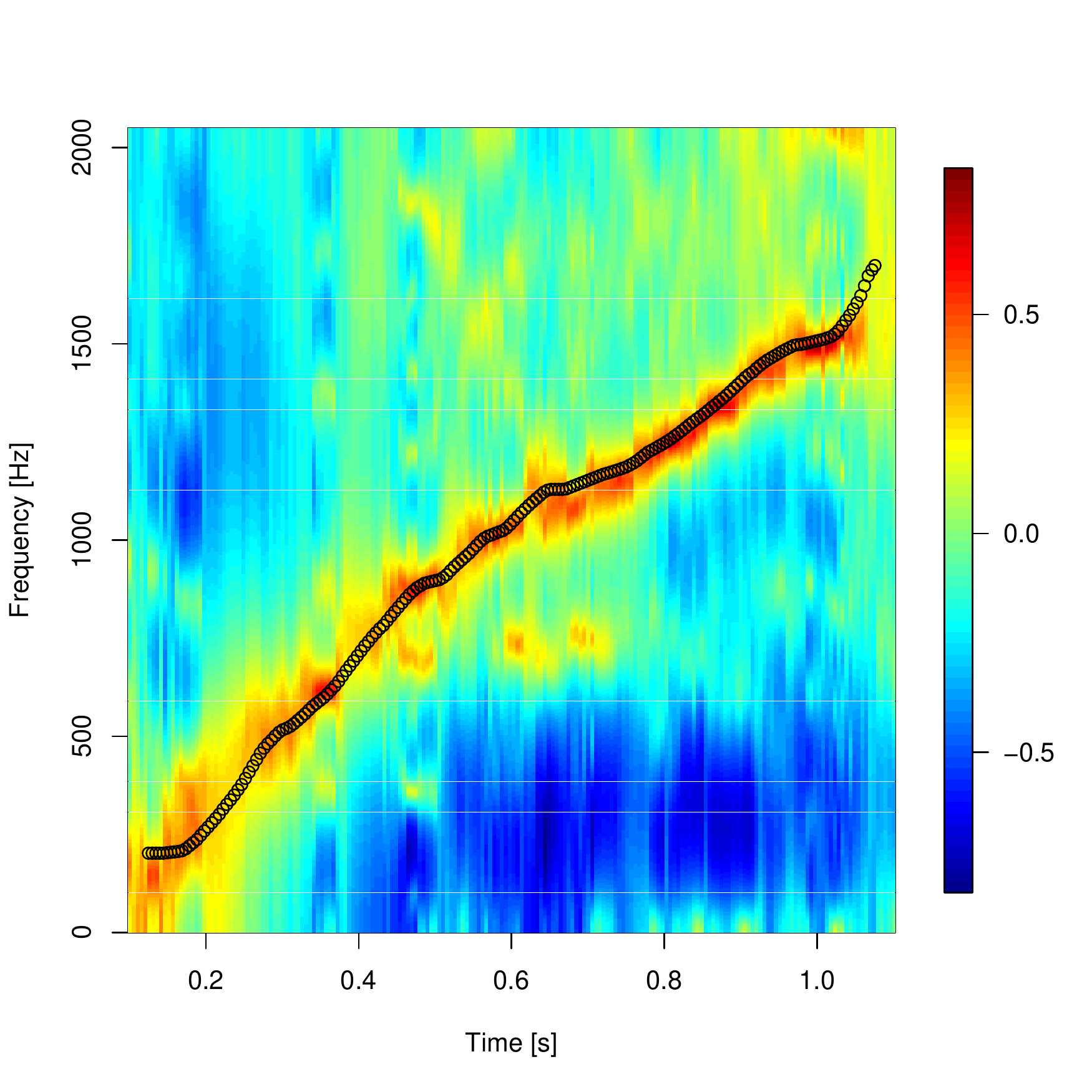}
 \caption{Spectrogram of the GW signal {\texttt s20S} sampled at \unit[4096]{Hz}.
   The spectrogram is obtained using a data streach of 200 samples overlapping at 90\%
   with each other. The open circles track the ridge $m(t) $ of the $\mbox{}^2 g_2$-mode. } \label{fig:spectrogram}
\end{figure}

We collect the instantaneous frequency $f(t_i)$ corresponding to the ridge $m(t_i)$ for
the midpoint $t_i$ of each local time interval of the spectrogram and interpolate $f(t)$
for values in between $t_i$. We then use our model given by Eq.~\eqref{eq:model1} to obtain
estimates of the time evolution of the ratio together with 95\% confidence intervals.
An example is given in Figure \ref{fig:ratio} where the red triangles are the point estimates and
the grey bands represent 95\% confidence bands. The size of the red triangles is proportional to the magnitude of the $\mbox{}^2 g_2$-mode frequency estimates.
Note that as the frequency of the $\mbox{}^2 g_2$-mode becomes higher our estimates show more uncertainty (bigger intervals) because our model allows for heterogeneous variance. Ratio values
computed using the mass and radius values obtained from the simulation code (i.e.~the true values)
are shown in black. In this example of a well detectable GW signal the coverage of our 95\% confidence band is 100\%
of the true values. In the next section we investigate the performance of the reconstruction of $r(t)$ when the GW
signal is embedded in noise. We also note that despite we have only explicitely shown results for the GW signal of the {\texttt s20S} model the same conclusions hold for any of the other waveforms of our test set.

\begin{figure}
 \centering
 \includegraphics[width=0.5\textwidth]{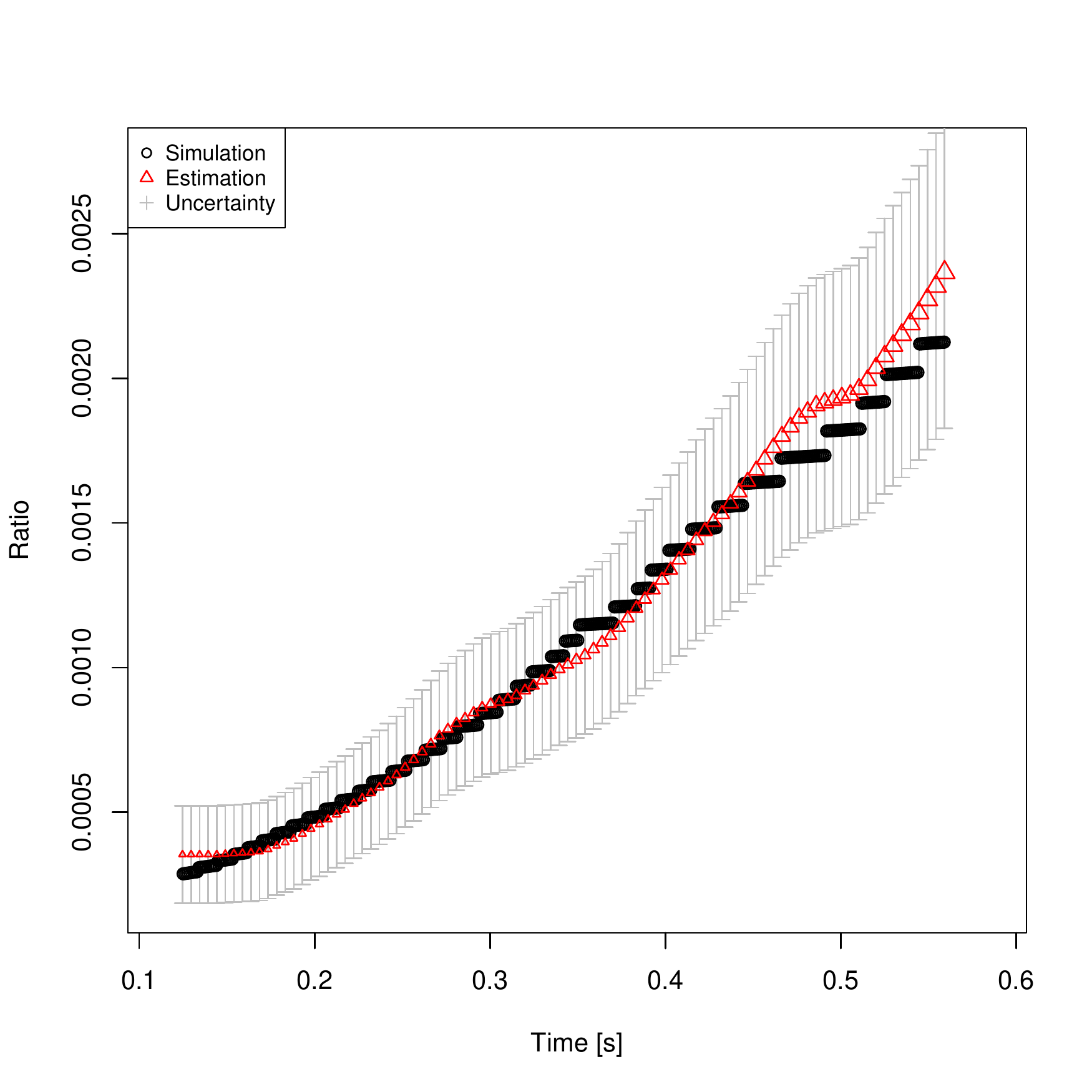}
 \caption{Comparison of the time evolution of the ratio $M_{\rm PNS}/R_{\rm PNS}^2$ estimated from the $\mbox{}^2 g_2$-mode of the {\texttt s20S} signal (shown by open triangles and by the 95\% confidence belt in grey) against the value derived from the PNS mass and radius given by the simulation code (shown by filled black circles). The size of the triangles are represented proportionally to the magnitude of the $\mbox{}^2 g_2$-mode frequency estimates.}
 \label{fig:ratio}
\end{figure}

\bigskip

% !TEX root = ccsn.tex
%\section{Detection sensitivity with Advanced gravitational wave detectors}
\section{Detectability prospects}
\label{sec:results}

%\begin{figure}[t]
% \centering
% \includegraphics[width=0.5\textwidth]{plots/spectrum}
% \caption{Amplitude spectral density of the GW detectors aLIGO and AdV at design sensitivity and that of the proposed third-generation detectors Cosmic Explorer and Einstein Telescope. Einstein Telescope sensitivity curve ET-B is obtained pushing second-generation detector technology at its limit. ET-C and ET-D sensitivity curves correspond to a detector configuration where a low-power cryogenic low-frequency interferometer and a high-power room temperature high-frequency interferometer are sharing the same infrastructure~\cite{Hild_2011}. Cosmic Explorer design sensitivity will be achieved in two stages. Stage 1 (CE1) is expected to use the technology developed for the ``A+'' upgrade to aLIGO but scaled up to a 40 km detector while stage 2 (CE2) will implement state-of-the-art technology to decrease quantum and thermal noises~\cite{reitze2019cosmic}.} 
% \label{fig:spectrum}
%\end{figure}

\begin{figure}[t]
  \centering
  \begin{tabular}{c}
    \includegraphics[width=0.5\textwidth]{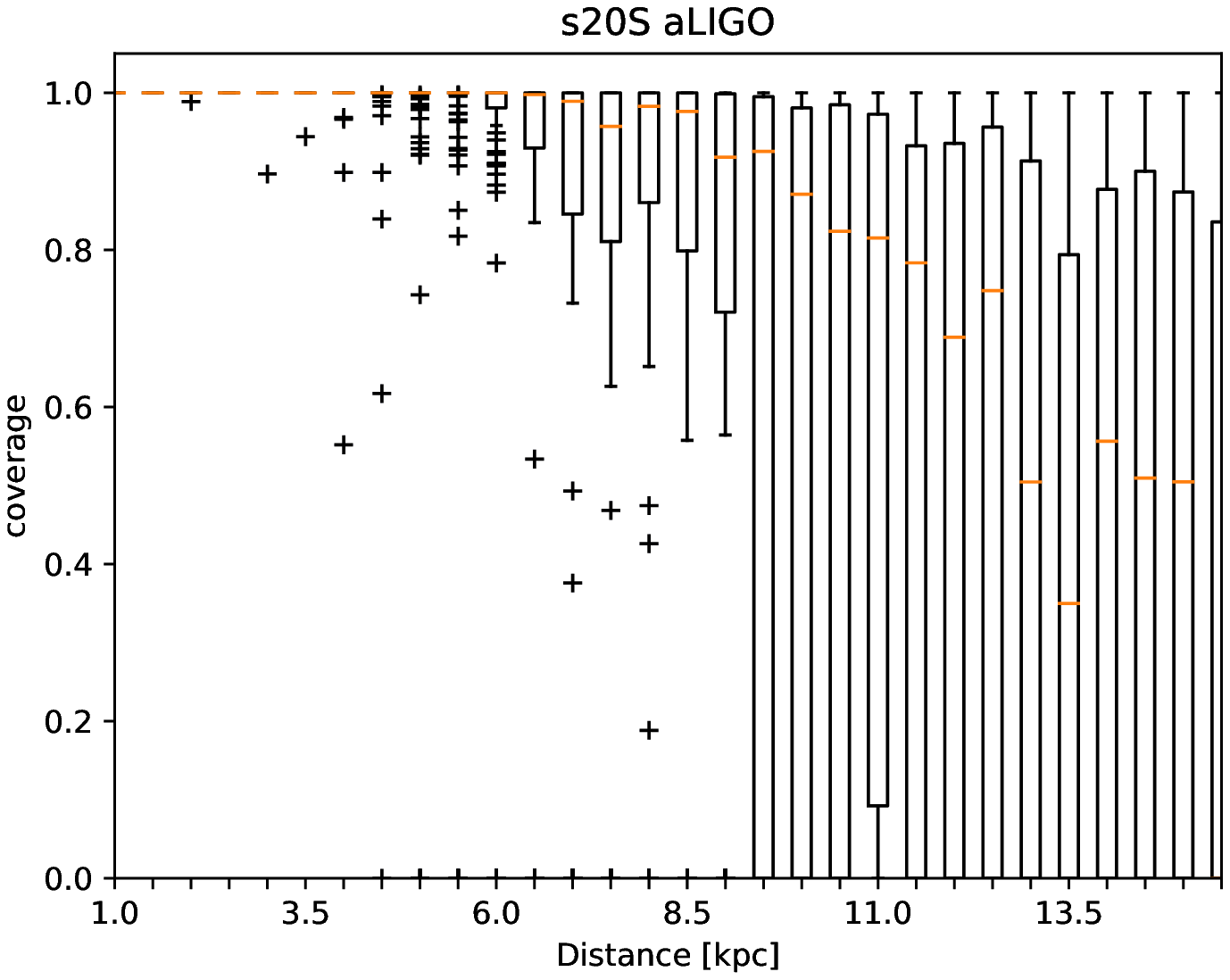} \\
    \includegraphics[width=0.5\textwidth]{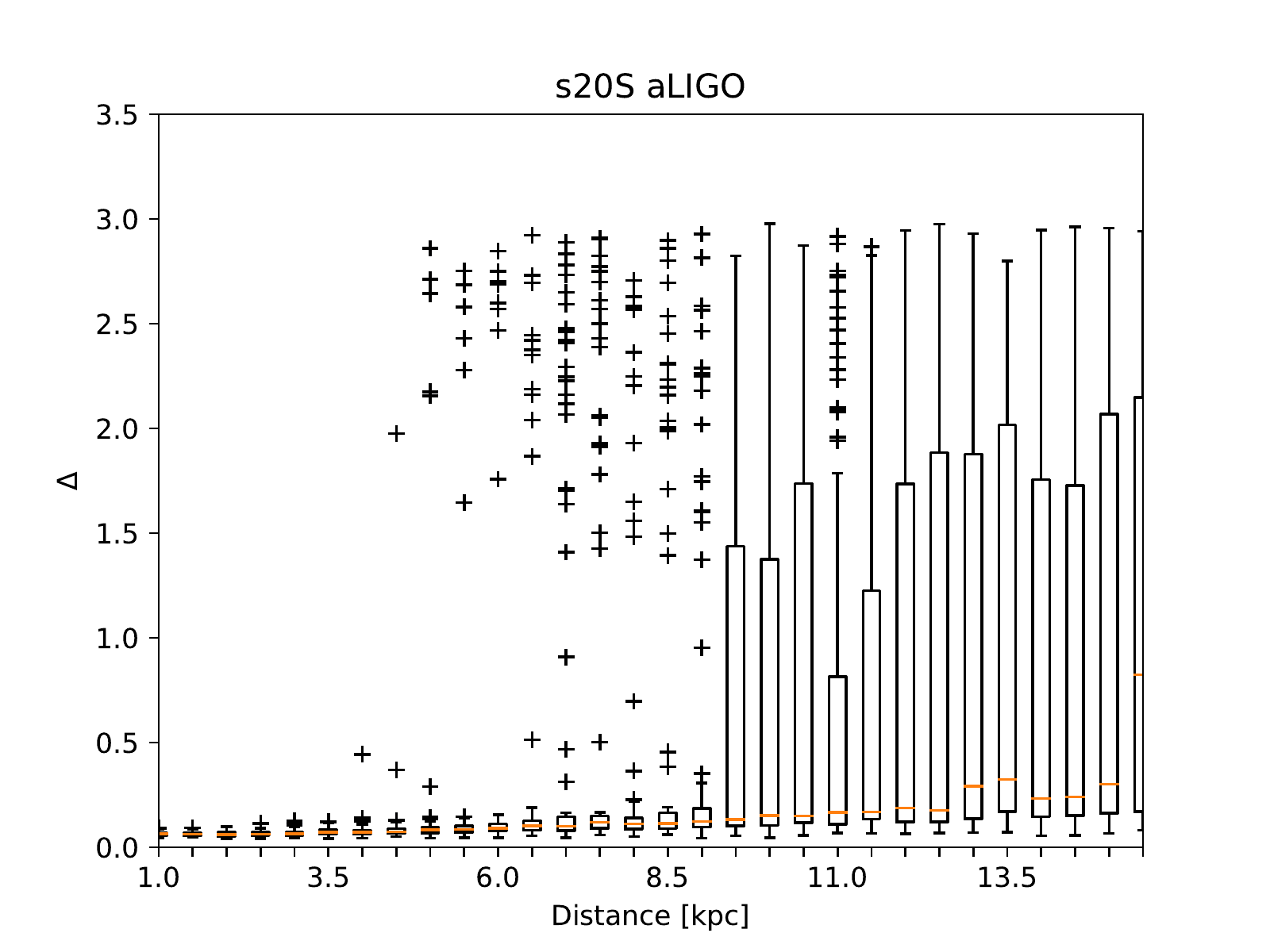} \\
  \end{tabular}
    
 \caption{Boxplots of the $coverage$ (upper panel) and $\Delta$ (lower panel) for {\texttt s20S} signal embedded in aLIGO noise at different distances from the Earth. 100 noise realizations are considered for each distance. The orange lines indicate the median values while the empty rectangles indicate the first and third quartiles. The ``+'' markers are outliers outside the first and third quartiles.}
  \label{fig:s20results}
\end{figure}

To estimate how accurately we can infer the time evolution of $r(t)$ in the GW data of a single
detector we inject the GW signal for model {\texttt s20S} into 
100 Gaussian noise realisations whose power spectral density (PSD) follows the aLIGO
spectrum~\citep{aLIGOsens:2018}. %The PSD of aLIGO and AdV at design sensitivities, along with those of planned third-generation detectors, are shown in Figure~\ref{fig:spectrum}. 

We cover a large range of distances for which a CCSN detection in second-generation GW detectors is feasible. We assume that the source is optimally oriented with respect to our single detector. Moreover, we also assume that a CCSN GW signal has been identified in the data and that the beginning of the signal is known within {$\mathcal O$}(10 ms). The data (signal embedded in noise) are whitened using the function {\tt prewhiten} of the R-package {\tt TSA}. An auto-regressive model with a maximum of maximal 100 coefficients is used.    

For each of the noise realizations we reconstruct the track ridge $m(t_i)$ starting from the left side of the spectrogram and constraining the beginning of the track to be smaller than \unit[200]{Hz}. This value is chosen using the information on the initial mode frequency from the simulations. We derive the ratio time series $r_i$ of length $N$ which is then compared to the ``true'' ratio {$r_i^0$} derived from the PNS mass and radius computed from the {\texttt s20S} simulation. The top panel of Fig.~\ref{fig:s20results} shows the distribution of the fraction of the ratio values {$r_i^0$} that fall within the 95\% confidence interval of {$r_i$} as a function of the distance to the source. This quantity, hence, gives information about the {\it coverage} of the reconstructed ratio. The coverage takes maximum values when the source is located within a few kpc and then decreases with the distance.

To better quantify how well we reconstruct the ratio we also consider  the mean of the relative error of $r_i$ along 
the track of the spectrogram, $\Delta$,  
\begin{equation}
\Delta=\frac{1}{N}\sum_1^N\frac{|r_i-r_i^0|}{r_i^0}\,.
\end{equation}
The values of $\Delta$ for each of the 100 noise realizations are shown as a function of the distance
in the bottom panel of Fig.~\ref{fig:s20results}. For a source located up to $\sim$\unit[9]{kpc} the relative error
remains smaller than 20\%. At closer distances $\Delta$ is smaller but it does not vanish. This reflects the approximate nature of the model used for $r$. It is nevertheless remarkable that, on average, one can reconstruct the ratio time series with a good
precision up to distances of $\sim$ \unit[9]{kpc} (for this particular waveform) with a coverage value
larger than 80\%. We note that there are a few noise realizations for distances below \unit[9]{kpc} for which
$\Delta$ takes large values, indicating that the method fails to accurately reconstruct the ratio in those cases. The main cause of failure is the split of the ridge in few blobs when the signal becomes weaker and weaker.
%\tf{Do we know why that is the case? What is intrinsically different from one noise realisation to another that makes the method more prone to failure? I think we should provide some explanation here, if we have it.}

\begin{figure}[t]
  \centering
  \includegraphics[width=0.5\textwidth]{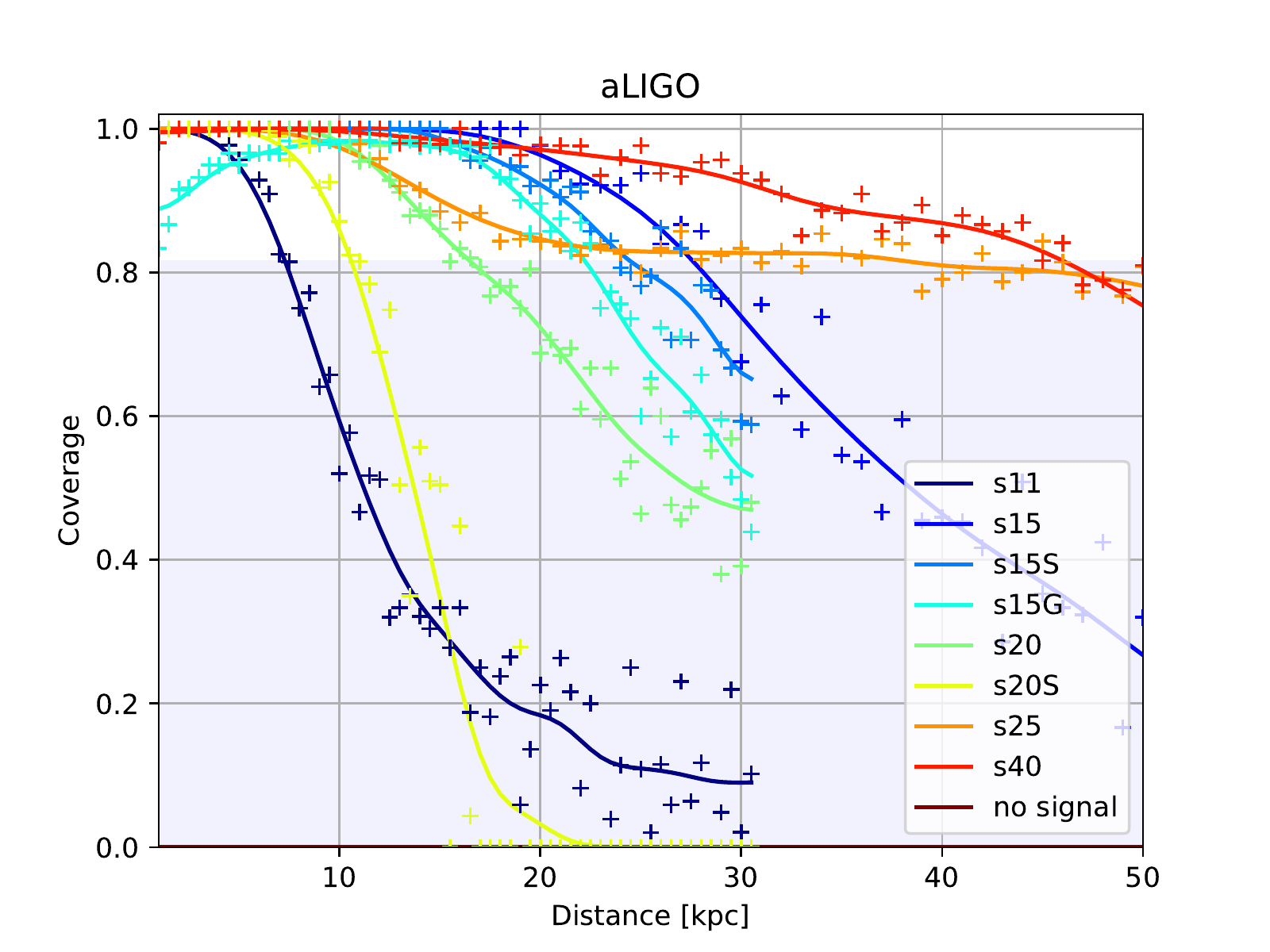}
 \caption{Data points (``+'') show the median of the coverage for the eight CCSN waveforms of the {\it test set} embedded in aLIGO noise as a function of the  distance to the source. The solid lines are smoothing splines. The ``no signal'' line shows the median of coverage in absence of any signal. In this example, the median is null and overlaps with the horizontal axis. The blue band boundaries are given by the $\mathrm{5^{th}}$ and $\mathrm{95^{th}}$ percentiles of coverage in absence of any signal.} \label{fig:aLIGO_cov_allwvf}
\end{figure}

\begin{figure}[t]
  \centering
  \includegraphics[width=0.5\textwidth]{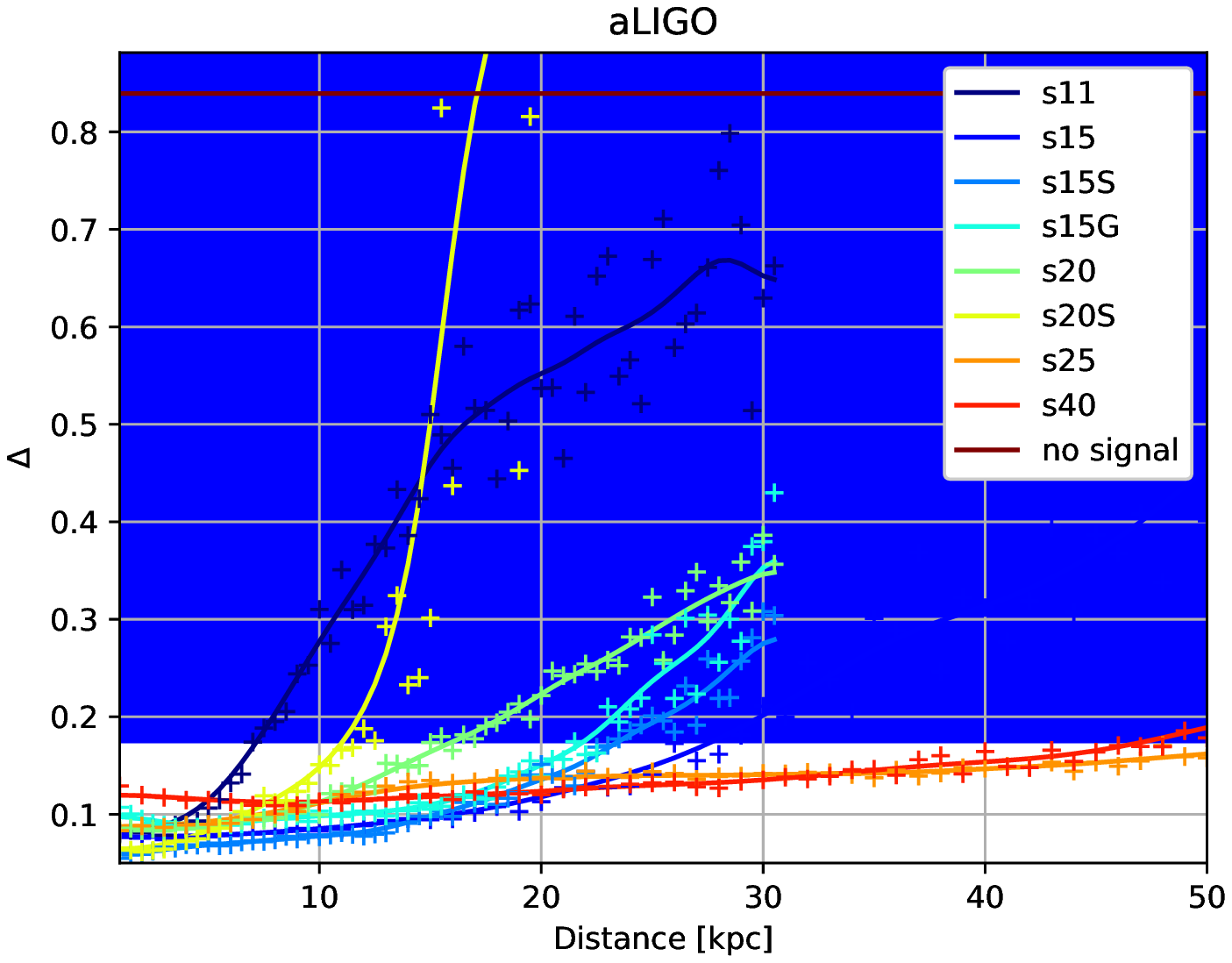}
  \caption{Data points (``+'') show the median of the relative error $\Delta$ for the eight CCSN waveforms of the {\it test set} embedded in aLIGO noise as a function of the distance to the source. The solid lines are smoothing splines. The ``no signal'' line shows the median of $\Delta$ in absence of any signal. The blue band boundaries are given by the $\mathrm{5^{th}}$ and $\mathrm{95^{th}}$ percentiles of $\Delta$ in absence of any signal.}\label{fig:aLIGO_prec_allwvf}
\end{figure}

We have tested that the method does not depend on the specific features of the waveform of model {\texttt s20S} by repeating the procedure for the remaining seven waveforms of the {\it test set} described in Section \ref{sec:simulations} covering
a large range of progenitor masses. Figure \ref{fig:aLIGO_cov_allwvf} shows that apart from model {\texttt s11} and to a lesser extent model {\texttt s20S}, the ratio is well reconstructed for all waveforms up to a distance of $\sim$ 15kpc. In an
effort to better determine the maximal distance of the source at which we can reconstruct the ratio we have run 100 simulations without injecting a signal and have measured the corresponding coverage for the reconstructed ratios.
The median of the coverage as well as the band defined by the $\mathrm{5^{th}}$ and $\mathrm{95^{th}}$ percentiles are shown in Figure \ref{fig:aLIGO_cov_allwvf}.
The noise only median value is identically zero in this case. However, note that it could be different
from zero because
the g-mode reconstruction algorithm is looking for a continuously  increasing frequency track
in the spectrogram, starting between 0 and 200 Hz, where we expect the GW signal to be.
This is enhancing the probability of overlap. This effect explains why certain values of overlap can reach
values as high as 80\% even when no signal is added to the noise.
%\tf{I don't understand this last comment.} \mab{Toni: is that clearer?}

Figure \ref{fig:aLIGO_prec_allwvf} shows the relative error $\Delta$ as a function of the distance for the signals of the {\it test set}
as well as the result when only noise is considered. This quantity follows the same trend than that followed by the coverage, since all signals but models {\texttt s11} and {\texttt s20S} are reconstructed with relative errors below 20\% up to distances of $\sim$ 15kpc. Correspondingly, the no-signal case yields the largest error, as expected.

We perform the same analysis using the design sensitivity curve of AdV and expected sensitivity curves for third-generation 
GW detectors. The results for the former are reported in Table \ref{tab:results} and are discussed below. We focus now on third-generation detectors, presenting our findings in Table \ref{tab:results} and in Fig.~\ref{fig:s20--SFHo_all3G}. In Europe the Einstein Telescope project proposes to host in a 10-km equilateral triangle configuration three low-power, low-frequency, cryogenic interferometers as well as three high-power, high-frequency interferometers. Three sensitivity curves, ET\_B, ET\_C and ET\_D corresponding to different options and stages of the project~\citep{Hild_2011} are considered in our study. %(cf.~Fig.~\ref{fig:spectrum}).
The US based project Cosmic Explorer~\citep{reitze2019cosmic} is expecting to reach its design
sensitivity circa 2040 through two phases labeled CE1 and CE2. %Their corresponding sensitivity curves are also shown in Fig.~\ref{fig:spectrum}. 

\begin{figure}[t]
  \centering
  \includegraphics[width=0.5\textwidth]{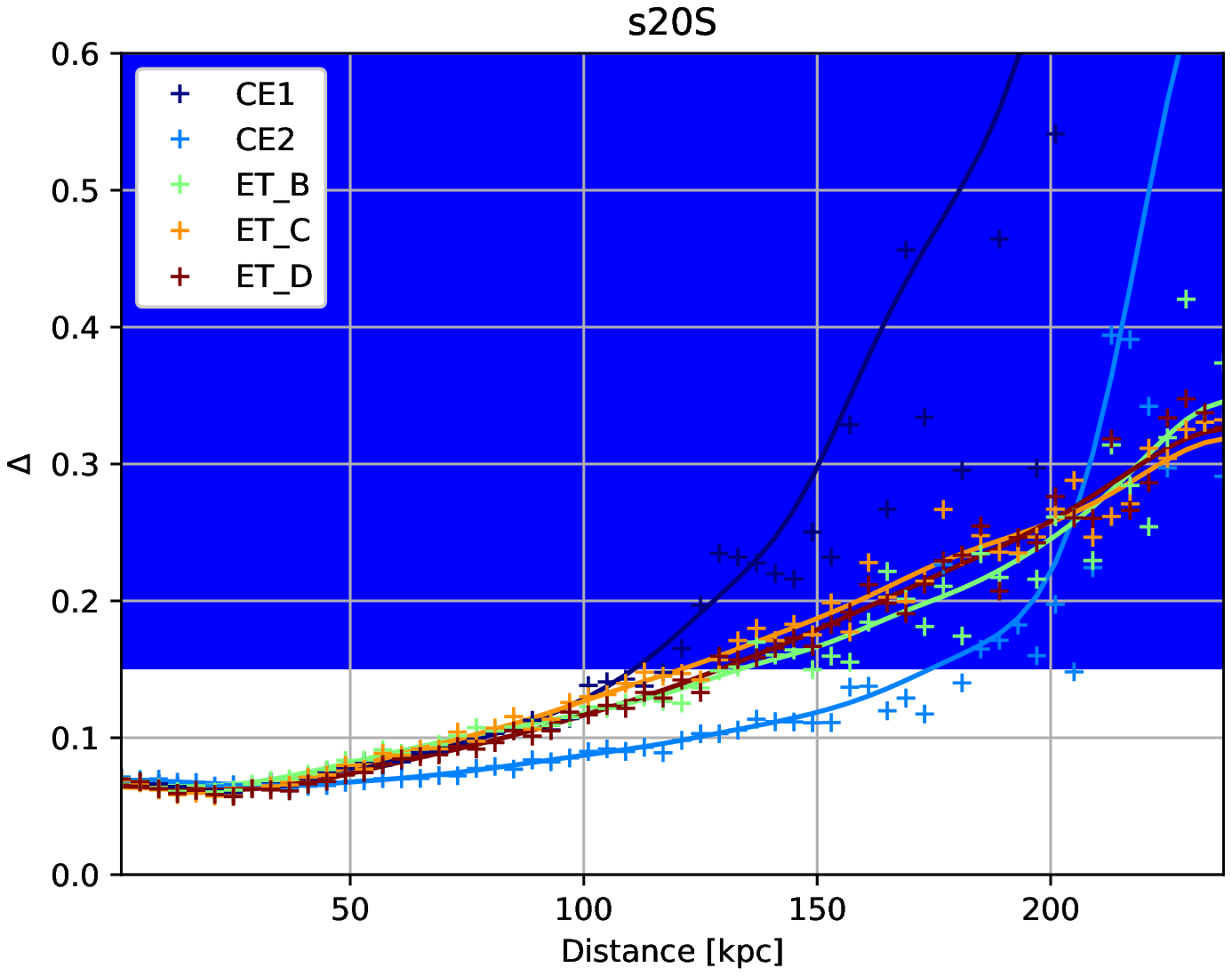}
  \caption{Data points (``+'') show the median of the relative error $\Delta$ as a function of the  distance to the source for the CCSN waveform model {\texttt s20S} embedded in third-generation detector noise. The solid lines are smoothing splines. The blue band boundaries are given by the $\mathrm{5^{th}}$ and $\mathrm{95^{th}}$ percentiles of coverage in absence of any signal for ET\_D noise realisations.}
  \label{fig:s20--SFHo_all3G}
\end{figure}

Figure \ref{fig:s20--SFHo_all3G} displays $\Delta$ as a function of the source distance for the five third-generation detector configurations we analyze. As before, we use the {\texttt s20S} waveform model as a reference case.  This figure shows that,  overall, the ratio is well reconstructed up to distances in the range \unit[100--200]{kpc} which represents an order of magnitude improvement with respect to aLIGO and AdV. We also note that the results for the various Einstein Telescope configurations lay in between those of the two Cosmic Explorer designs. Moreover,  the detectability prospects for the former depend weakly on the detector configuration while the arrangement of CE2 yields better results than CE1. These results are confirmed for all other waveforms of our sample except for model {\texttt s25} for which the maximal distance reach in CE2 is significantly lower than CE1, as shown in Fig.~\ref{fig:distances}. This is partly due to the small variation of the reconstruction quality to the distance of the source making the estimation of $d_r$ rather uncertain for this particular waveform. %\tf{Is this really understood? What's particular in this waveform?}

\begin{figure}[t]
  \centering
  \includegraphics[width=0.5\textwidth]{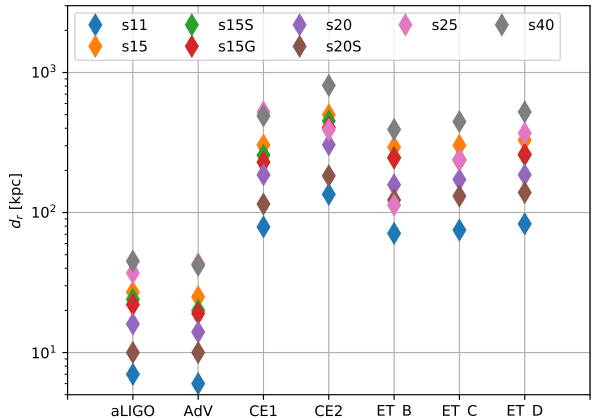}
  \caption{Maximal distance at which the ratio $r=M_{\rm PNS}/R_{\rm PNS}^2$ is reconstructed
    with good accuracy for all GW detectors analyzed in this study. The values are shown for all eight CCSN waveforms of our sample assuming that the source is optimally oriented with respect to the detector. Some waveform maximal distance markers are not visible because the do overlap (for instance \texttt{s15G} is overlapping with \texttt{s25} for CE2).} 
\label{fig:distances}
\end{figure}

\begin{table}
  \centering
  \begin{tabular}{c|c|cccccccc}

\multicolumn{2}{c|}{}  & \texttt{s11} & \texttt{s15} & \texttt{s15S} & \texttt{s15G} & \texttt{s20} & \texttt{s20S} & \texttt{s25}  & \texttt{s40}\\   

\hline
\multirow{3}{*}{aLIGO} & $d_{r}$   & 7 & 28 & 24  & 22 & 16 & 11 & 38 & 46 \\
\cline{2-10}
                       & $d_{\rm det}$ & 11 & 36 & 26 & 27 & 21 & 16 & 74 & 61\\
%                      & SNR      & 14 & 46 & 33 & 35 & 27 & 21 & 96 & 80\\

\hline
\hline
\multirow{3}{*}{AdV}   & $d_{r}$   & 7  & 26  & 20 & 19 & 15 & 10 & 43 & 42 \\
\cline{2-10}
                       & $d_{\rm det}$ &  10 & 32 & 22 & 23 & 18 & 13 & 64 & 52\\
%                      & SNR      &  13 & 41 & 29 & 30 & 23 & 17 &  83 & 68\\

\hline
\hline
\multirow{2}{*}{CE1}   & $d_{r}$   & 79  & 304 & 258 & 229 & 187 & 115 & 524 & 490 \\
\cline{2-10}
                       & $d_{\rm det}$ & 115 & 377 & 270 & 282 & 217 & 168 & 774  & 633\\
%                      & SNR      & 149 & 490 & 352 & 366 & 282 & 218 & 1006 & 822\\

\hline
\multirow{2}{*}{CE2}   & $d_{r}$  & 135 & 499 & 451 & 405 & 305 & 183 & 391 & 898 \\
\cline{2-10}
                       & $d_{\rm det}$ & 197 & 649 & 468 & 489 & 375 & 294 & 1347  & 1100\\
%                      & SNR    & 256 & 843 & 608 & 635 & 487 & 382 & 1751 & 1430\\

\hline
\multirow{2}{*}{ET\_B} & $d_{r}$ & 71  & 293 & 248 & 245 & 158 & 123 & 113 & 392 \\
\cline{2-10}
                       & $d_{\rm det}$ & 106 & 364 & 274 & 391 & 216 & 200 & 805 & 665\\
%                      & SNR      & 138 & 473 & 356 & 379 & 381 & 260 & 1046 & 865\\

\hline
\multirow{2}{*}{ET\_C} & $d_{r}$ & 75  & 302 & 239 & 237 & 172 & 131 & 239 & 446 \\
\cline{2-10}
                       & $d_{\rm det}$ & 97 & 332 & 246 & 260 & 194 & 164 & 727  & 603\\
%                      & SNR      & 126 & 432 & 320 & 338 & 252 & 213 & 945 & 783\\

\hline
\multirow{2}{*}{ET\_D} & $d_{r}$ & 83  & 329 & 257 & 261 & 186 & 139 & 369 & 523 \\
\cline{2-10}
                       & $d_{\rm det}$ & 107 & 368 & 271 & 285 & 213 & 174 & 796  & 661\\
%                      & SNR      & 140 & 477 & 352 & 371 & 277 & 227 & 1034 & 859 \\

  \end{tabular}
  \caption{%%
  Maximal distance $d_{r}$ at which the ratio $r=M_{\rm PNS}/R_{\rm PNS}^2$ is reconstructed
    with good accuracy for all GW detectors analyzed in this study, assuming optimal orientation between the source and the detector. Correspondingly, $d_{\rm det}$ is the distance at which different interferometers could detect a source
    optimally oriented with a matched filter signal-to-noise ratio of 13. All distances are expressed in kpc.
  }
  \label{tab:results}
\end{table}

All results for both second-generation and third-generation detectors are summarized in Table \ref{tab:results} and Figure
\ref{fig:distances}. Table \ref{tab:results} reports the source distances $d_r$ in kpc at which the median of the coverage is lower than 95\% of the noise only values for aLIGO, AdV, and different configurations of third-generation detectors. This same information is displayed in Fig.~\ref{fig:distances}. We have checked that using either the median of the coverage or the median of $\Delta$ yields similar results for the distance. On the other hand, the quality of the ratio reconstruction and, thus, of the distance range, depends on the signal-to-noise ratio, expressed in Table \ref{tab:results} by $d_{\rm det}$. The numbers reported on the table for $d_r$ are an estimate of the order of magnitude of the maximal distance of the source  at which a
reconstruction of the ratio could be possible with current and planned GW detectors. We also provide upper limits for $d_{\rm det}$ by taking into account the detector antenna response in our simulations and assuming that the source is optimally oriented with a matched filter signal-to-noise ratio of 13. Table \ref{tab:results} shows that the results for the AdV detector at design sensitivity are very similar to those of aLIGO, despite the differences in detector sensitivity. It is remarkable that for third-generation detectors the PNS surface gravity  could be reconstructed for sources located up to a few hundreds of kpc. It is nevertheless important to note the rather wide range in distances we obtain for the different waveforms of our {\it test set} that probe a large range of progenitor masses. We do not find any correlation between either the mass of the progenitor nor the EOS with $d_r$.

\bigskip

% !TEX root = ccsn.tex
\section{Conclusions}
\label{sec:conclusion}

The detection of GW from CCSN may help improve our current understanding of the explosion mechanism of massive stars.  
In this paper we have proposed an exploratory method to infer PNS properties using an approach based on GW associated with convective oscillations of PNS. As shown by~\citep{Torres:2019b} buoyancy-driven g-modes are excited in numerical simulations of CCSN and their time-frequency evolution is linked to the physical properties of the compact remnant through universal relations. Such modes are responsible for a significant fraction of the highly stochastic GW emitted after core bounce. The findings reported in this paper suggest that PNS asteroseismology might be within reach of current and third-generation GW detectors.

In our study we have used a set of 1D CCSN simulations to build a model that relates the  evolution of PNS properties with the frequency of the dominant g-mode, namely the $\mbox{}^2g_2$ mode. This relationship is extracted from the GW data using an algorithm developed for this investigation. This algorithm is  a first attempt to infer the time evolution of a particular combination of the PNS mass and radius based on the universal relations found in~\citep{Torres:2019b}. More precisely, we have considered the ratio $r=M_{\rm PNS}/R_{\rm PNS}^2$ (the PNS surface gravity) derived from the observation of the $\mbox{}^2g_2$ oscillation mode in the numerically generated GW data. The  performance of our method has been estimated employing  simulations of 2D CCSN waveforms covering a progenitor mass range between 11 and 40 solar masses and different equations of state. 

We have investigated the performance of the algorithm in the case of an optimally oriented source detected by a singe GW detector. Our numerical signals have been injected into 100 Gaussian noise realisations whose PSD follow the spectra of the different GW detectors analyzed. We have found that for Advanced LIGO and Advanced Virgo, the ratio $r$ can be reconstructed with a good accuracy in the case of a galactic CCSN (i.e.~for distances of ${\cal O}(10\, {\rm kpc})$). This holds for a wide range of progenitor masses, the quality of the inference mainly depending on the signal-to-noise ratio of the event. For third-generation GW detectors such as the Einstein Telescope and the Cosmic Explorer, however, we obtain an order of magnitude improvement, as the $\mbox{}^2g_2$ ratio can be reconstructed for sources at distances of ${\cal O}(\unit[100]{kpc})$. In particular, Cosmic Explorer in its stage 2 configuration yields the best performance for all waveforms we have considered thanks to its excellent sensitivity in the \unit[100-1000]{Hz} range. Among the three configurations of the Einstein Telescope, ET-D provides the best performance, especially for our set waveforms with the highest progenitor masses (25 $\Msol$ and 40 $\Msol$). Comparing the estimated distances for ET-B and the other third-generations detectors, having a good sensitivity below \unit[200]{Hz} seems the most important factor to detect high mass progenitor signals.

In the present study we have assumed that the sources are optimally oriented. The reported distances at which
we can infer the time evolution of $M_{\rm PNS}/R_{\rm PNS}^2$ must thus be regarded as upper limits. Those figures may decrease by a factor 2--3 on average for a source located anywhere in the sky. Furthermore, we have not considered the detectability prospects of  CCSN waveforms in the realistic case in which the  interferometers operated as a detector network. 
We defer an improved implementation of our approach for a forthcoming publication. Finally, we note that the method discussed in this work can be adapted to other PNS oscillation modes, by simply changing a few parameters such as the initial frequency range of the mode and its monotonic raise or descent. Being able to reconstruct several modes in the same GW signal would potentially allow to individually infer the mass and the radius of the PNS in core-collapse supernova explosions.

%Furthermore, the results presented here are based on a model of the frequency evolution of the
%ratio that is making use of AENUS-ALCAR code simulations. Using simulations produced
%with the CoCoNut code, the model fit is a bit different (see Figure \ref{fig:LMVAR}). This systematic
%difference of the two simulation codes is discussed in \cite{Torres:2019b}. If we consider a model based
%on CoCoNut simulations and test the method on test simulations obtained with the AENUS-ALACR code,
%we observe systematic effects in the ratio reconstruction leading to systematically worsen performance.
%As the goal of this paper is to characterize the methods and because both simulation codes have pros
%and cons we have reported results obtained with a model set and test set waveforms obtained with the same
%code.

\bigskip

\bigskip\noindent\textit{Acknowledgments.--} Work supported by the Spanish Agencia Estatal de Investigaci\'on (PGC2018-095984-B-I00), by the Generalitat Valenciana (PROMETEO/2019/071) and by the European Union’s Horizon 2020 research and innovation (RISE) programme (H2020-MSCA-RISE-2017 GrantNo. FunFiCO-777740).
N.C. acknowledges support from National Science Foundation grant PHY-1806990.
R.M. gratefully acknowledges support by the James Cook Fellowship from Government funding, administered by the Royal Society Te  Ap\={a}rangi. R.M. and P.M.-R. also acknowledge funding from  DFG Grant KI 1443/3-2 and thank the Centre for eResearch at the University of Auckland for their technical support.
P.C.-D. acknowledges the support of the Spanish {\it Ramon y Cajal} programme (RYC-2015-19074) supporting his research.
\begin{appendices}
\section{g-mode reconstruction}
\label{app:gmode}
Given the spectrogram and a prescribed time interval for the $^2g_2$-mode reconstruction, our proposed method works as follows.  The starting point must be specified.  It can be either at the beginning or at the end of the signal.  Then, in one of these two extremes, the maximum energy value is identified, registering its frequency.  This is done independently for a number of consecutive time intervals.  Then, we calculate the median of these frequency values, providing a robust starting value for the $^2g_2$-mode reconstruction.

The starting frequency value is the first $^2g_2$-mode estimate for the first or the last time interval, depending on the starting location we choose.  If the reconstruction is set to start at the beginning of the signal, the reconstruction will be done progressively over the time intervals, where each maximum frequency value will be calculated within a frequency range specified by the previous $^2g_2$-mode estimate.  Given the non-decreasing behaviour of the true $^2g_2$-mode values, the mode estimates will be forced to be greater or equal than the one estimated for its previous time interval, and lower than a specified upper limit.  As a result, the $^2g_2$-mode estimates will be a non-decreasing sequence of frequency values. Then, the moving average is applied for smoothing the estimates.

If the reconstruction is set to start at the end of the signal, the $^2g_2$-mode will be estimated backward in time.  Each maximum frequency is calculated within a range determined by its successor (in time) mode estimate.  These estimates are forced to be lower or equal than its successor (in time) estimate, but greater than a specified lower limit. Thus, a non-decreasing sequence of $^2g_2$-mode estimates is guaranteed.  Then, the moving average is applied for smoothing the estimates. This $^2g_2$-mode reconstruction method works if and only if the signal is strong enough to provide information about the mode, which is reflected in the spectrogram.

Given the sequence of $^2g_2$-mode estimates, the confidence band will be calculated by using the model defined in Eq.~\eqref{eq:model1}. The $^2g_2$-mode estimates are frequency values which we use as predictors in the model in order to generate confidence intervals for the ratios. Since the mode estimates are indexed by time, the confidence intervals for the ratios are too.  Thus, we generate the confidence band by interpolating the lower and upper limits of the collection of consecutive confidence intervals, which will be valid for the time range of the $^2g_2$-mode estimates.  This confidence band is used to estimate the coverage probabilities in our simulation studies presented in the main text.  

\end{appendices}
\bibliography{biblio}

\end{document}